\documentclass[preprint,amsmath,amssymb,aps,floatfix,groupedaddress,superscriptaddress,nofootinbib,preprintnumbers]{revtex4-1}
\usepackage{float}
\usepackage{amsmath}
\usepackage[pdftex]{hyperref}% add hypertext capabilities
\usepackage{graphicx,color}% Include figure files
\usepackage{dcolumn}% Align table columns on decimal point
\usepackage{bm}% bold math
\usepackage{here}
\usepackage{comment}
\usepackage[paperwidth=210mm,paperheight=297mm,centering,hmargin=1.5cm,vmargin=2.0cm]{geometry}
\usepackage{tabularx}

\allowdisplaybreaks[1]

\makeatletter
\newsavebox{\@brx}
\newcommand{\llangle}[1][]{\savebox{\@brx}{\(\m@th{#1\langle}\)}%
  \mathopen{\copy\@brx\mkern2mu\kern-0.9\wd\@brx\usebox{\@brx}}}
\newcommand{\rrangle}[1][]{\savebox{\@brx}{\(\m@th{#1\rangle}\)}%
  \mathclose{\copy\@brx\mkern2mu\kern-0.9\wd\@brx\usebox{\@brx}}}
\makeatother

\begin{document}

\title{Pileup Correction on Higher-order Cumulants with Unfolding Approach} 

\author{Yu Zhang}
\email{zhang\_yu@mails.ccnu.edu.cn}
\affiliation{Key\,Laboratory\,of\,Quark\,\&\,Lepton\,Physics\,(MOE)\,and\,Institute\,of\,Particle\,Physics,\,Central\,China\,Normal\,University,\,Wuhan\,430079,\,China}
\affiliation{Lawrence Berkeley National Laboratory, CA 94720, USA}
\author{Yige Huang}
\email{yghuang@mails.ccnu.edu.cn}
\affiliation{Key\,Laboratory\,of\,Quark\,\&\,Lepton\,Physics\,(MOE)\,and\,Institute\,of\,Particle\,Physics,\,Central\,China\,Normal\,University,\,Wuhan\,430079,\,China}
%\item \href{https://orcid.org/0000-0002-2714-2119}{\textcolor{orcidlogocol}{\aiOrcid} \hspace{2mm} orcid.org/0000-0002-2714-2119}
\author{Toshihiro Nonaka}
\email{nonaka.toshihiro.ge@u.tsukuba.ac.jp}
\affiliation{Tomonaga\,Center\,for\,the\,History\,of\,the\,Universe,\,University\,of\,Tsukuba,\,Tsukuba,\,Ibaraki\,305,\,Japan}
\author{Xiaofeng Luo}
\email{xfluo@ccnu.edu.cn} 

\affiliation{Key\,Laboratory\,of\,Quark\,\&\,Lepton\,Physics\,(MOE)\,and\,Institute\,of\,Particle\,Physics,\,Central\,China\,Normal\,University,\,Wuhan\,430079,\,China}

%%%%%%%%%%%%%%%%%%%%%%%%%%%%%%%%%%%%%%%%

\begin{abstract}
	Higher-order cumulants of conserved charge distributions are sensitive observables to probe the critical fluctuations near QCD critical point in heavy-ion collisions. Due to high interaction rate, pileup event can be one of the major sources of background in the measurements of higher-order cumulants. In this paper, we studied the effects of pileup events on higher-order cumulants of proton multiplicity distributions using UrQMD model. It is found that the proposed pileup correction fails if the correction parameters are determined by the Glauber fitting of charged particle multiplicities, which is usually done in the real heavy-ion experiment. To address  this, we propose a model independent unfolding approach to determine the parameters in the pileup correction. This approach can be applied in the pileup correction for the future measurement of higher-order cumulants in heavy-ion collision experiment. 
\end{abstract}
\maketitle

%%%%%%%%%%%%%%%%%%%%%%%%%%%%%%%%%%%%%%%%
\newcommand{\ave}[1]{\ensuremath{\langle#1\rangle} }
\newcommand{\avebig}[1]{\ensuremath{\Bigl\langle#1\Bigr\rangle} }
\newcommand{\aveave}[1]{\ensuremath{\langle\!\langle#1\rangle\!\rangle} }

\onecolumngrid

\section{Introduction}
One of the main goals of heavy-ion collision experiments is to 
understand the phase structure of QCD matter with respect to baryon chemical 
potential ($\mu_{\rm B}$) and temperature ($T$)~\cite{Gupta:2011wh,Luo:2017faz,Bzdak:2019pkr}. 
The most important question is the position of a QCD critical 
end point, which is a connecting point between crossover and first-order phase transition, 
in ($\mu_{\rm B}$, T) plane~\cite{Stephanov:2004wx,Fukushima:2010bq,Shi:2014zpa,Ding:2015ona,Gao:2016qkh,Bazavov:2017dus,Fischer:2018sdj,Fu:2019hdw,Isserstedt:2019pgx,Gao:2020qsj,Bernhardt:2021iql,Borsanyi:2021hbk}.
In order to explore the QCD phase diagram, higher-order fluctuations of net-baryon, net-charge, and net-strangeness multiplicity distributions have been proposed~\cite{Hatta:2003wn,Ejiri:2005wq,Stephanov:2008qz,Asakawa:2009aj,Friman:2011pf} and  measured in ALICE~\cite{Arslandok:2020mda}, HADES~\cite{Adamczewski-Musch:2020slf}, NA61/SHINE~\cite{Mackowiak-Pawlowska:2020glz}, and STAR collaborations~\cite{Aggarwal:2010wy,net_charge,net_proton,Adamczyk:2017wsl,Adam:2020unf,Nonaka:2020crv,Abdallah:2021zhr,Luo:2015doi}.
Recently the STAR experiment reported a non-monotonic beam energy dependence of the fourth-order fluctuations of net-proton multiplicity distributions from the Beam Energy Scan program (BES-I) at RHIC~\cite{Adam:2020unf,STAR:2021iop}. This is consistent with the expectations of critical fluctuations in QCD-inspired models
~\cite{Schaefer:2011ex,Chen:2015dra,Shao:2017yzv,Fan:2017mrk,Li:2018ygx,Mroczek:2020rpm,Fu:2021oaw}.
More definitive physics messages will be achieved by enhanced statistics in Beam Energy Scan program phase II (BES-II, 2019-2021)~\cite{bes2}. According to the model calculations, the value of the fourth-order fluctuations decreases again to the statistical baseline at further low collision energies~\cite{Stephanov:2011pb}.
To confirm this the STAR experiment has been carrying out the fixed-target experiment for $3.0\leq\sqrt{s_{\rm NN}}\;\leq7.7$ GeV.

At low collision energies, the fixed-target experiment is more efficient than collider experiment due to much less interaction rate for the latter~\cite{Luo:2020pef}. 
However, the high interaction rate in the fixed-target experiment increases the probability of pileup collisions.
Pileup is a superposition of two or more single-collision events occurring within a short time interval. 
It is straightforward to correct a mean value of an observable for the pileup effects using known value of the pileup probability. 
Unfortunately, this is not the case for higher-order fluctuations.
A correction method on higher-order fluctuations was proposed in Ref.~\cite{Nonaka:2020qcm}.
The pileup correction was tested in simple toy model with pre-defined correction parameters. It was thus a closure test where the proposed method should always work perfectly.
In real experiments, however, correction parameters are unknown and need to be determined by simulations. The accuracy of the pileup correction strongly depends on how precisely the simulations can reflect the real conditions of the experimental data. To address this, we propose a model independent way of an unfolding approach to determine the correction parameters. We study the effect of pileup events on higher-order cumulants in UrQMD model. Pileup corrections are carried out in both closure test and Glauber model fits. These results are shown as a function of the pileup probability and compared to true values in UrQMD. Finally, the unfolding approach is applied to UrQMD samples for pileup corrections.

This paper is organized as follows. In Sec.~\ref{sec:pileup}, we introduce the methodology of pileup corrections on cumulants.
Pileup corrections are applied to UrQMD model in Sec.~\ref{sec:urqmd} 
In Sec.~\ref{sec:unfolding} we propose an unfolding approach to extract the 
correction parameters in a model independent way. The validity of the method is checked in UrQMD model. We then summarize the present study in Sec.~\ref{sec:summary}

%%%%%%%%%%%%%%%%%%%%%%%%%%%%%%%%%%%%%%%%%%%%%%%%%%%%%
\section{Moments and cumulants}
Let us consider a probability distribution of particle number P(N). 
The $r$th order moment is defined by
\begin{equation}
    \ave{N^{r}}=\sum_{N}N^{r}P(N),
\end{equation}
where the bracket represents the event average.
It is convenient to introduce a moment generating function,
\begin{equation}
G(\theta)=\sum_{N}e^{N\theta}P(N)=\ave{e^{N\theta}}.
\end{equation}
Then the $r$th order moment can be expressed as $r$th derivative of $G(\theta)$
\begin{equation}
    \ave{N^{r}}=\frac{d^{r}}{d\theta^{r}}G(\theta)\Bigl|_{\theta=0}.
\end{equation}
Since moments drastically increase with increasing the order $r$, cumulants are easier 
to handle rather than moments. 
A cumulant generating function is defined as
\begin{equation}
    K(\theta)={\rm ln}G(\theta).
\end{equation}
The $r$th order cumulant is then given by
\begin{equation}
    C_{r}=\frac{d^{r}}{d\theta^{r}}K(\theta)\Bigl|_{\theta=0}.
\end{equation}
In the rest of paper, moments will mainly appear for the pileup correction, while 
results will be discussed based on cumulants.

%%%%%%%%%%%%%%%%%%%%%%%%%%%%%%%%%%%%%%%%%%%%%%%%%%%%%
\section{Pileup correction\label{sec:pileup}}
%Experimentally, there is finite probability that more than one collision 
%events overlap each other, hich is called pileup events. 
%The source of pileup events are classified into out-of-bunch and 
%in-bunch pileups. The former one would be easily rejected depending on how precisely 
%the collision vertices are determined. 
%However, once more than one collision events occur within short time interval, 
%those collisions cannot be separated into single collision events experimentally. 
%This could be partially removed by offline analysis by looking at 
%the correlation between fast and slow detectors, but it is difficult to remove those events completely. 
%This effect will be more significant in fixed-target mode experiments rather than collider mode experiments, since the collision rate is much higher for the former one.

%According to the previous studies~\cite{Garg:2017agr,Sombun:2017bxi}, 
%higher-order cumulants get affected by pileup events, especially in 
%the most central collisions, but any correction method was not available. 
%In order to solve this problem, a method of pileup correction for higher-order cumulants was developed 
%in Ref.~\cite{Nonaka:2020qcm}.
Let $P_{m}(N)$ be a probability distribution function to find one event with $N$ particles at reference multiplicity $m$. Throughout this paper we suppose that pileup events are formed by independent superposition of two single-collision events with the probability $\alpha$. Then $P_{m}(N)$ can be rewritten as 
\begin{eqnarray}
P_{m}(N) = (1-\alpha_{m})P_{m}^{\mathrm{t}}(N) + \alpha_{m}P^{\mathrm{pu}}_{m}(N).	
\end{eqnarray} where $P_{m}^{\mathrm{t}}(N)$ and $P_{m}^{\mathrm{pu}}(N)$ are probability distribution functions for single collision event and pileup event respectively. Pileup events at multiplicity $m$  can be decomposed into sub-pileup events whose multiplicity satisfies $m$ = $i$ + $j$.
By looping all possible combinations of $i$ and $j$ which satisfies $m$ = $i$ + $j$, probability distribution function for pileup events is obtained and written as  
\begin{eqnarray}
	P_{m}^{\mathrm{pu}}(N) = \sum_{i,j}\delta_{m,i+j}w_{i,j}P_{i,j}^{\mathrm{sub}}(N),
\end{eqnarray}
%\begin{eqnarray}
%	P_{i+j}^{sub}(N) = \sum_{N_{i},N_{j}}\sigma_{N,N_{i}+N_{j}}P^{t}_{i}(N_{i})P_{j}^{t}(N_{j}),
%\end{eqnarray} 
where $w_{i,j}$ is the probability to observe a sub-pileup event among all pileup events at multiplicity $m$. $i$ and $j$ commutes in $w_{i,j}$ which gives $w_{i,j}$ = $w_{j,i}$. Exhausting all combinations of $i$ and $j$ there should be 
\begin{eqnarray}
\sum_{i,j}\delta_{m,i+j}w_{i,j}=1.	
\end{eqnarray}
 
We also consider a multiplicity distribution $T(m)$ used for centrality determination.
The pileup events at the $m$th multiplicity bin 
are then decomposed into two sub-pileup events which satisfies $m=i+j$.
%the fraction of pileup events at the $m$th multiplicity bin,$\alpha_{m}$
%Let us then introduce following notations:
Then we get~\cite{Nonaka:2020qcm}
\begin{eqnarray}
 w_{i,j} &=& \frac{\alpha T(i)T(j)}{\sum_{i,j}\delta_{m,i+j}\alpha T(i)T(j)},
  \label{eq:w_ij}
  \\
    \alpha_{m} &=& \frac{ \alpha \sum_{i,j}\delta_{m,i+j}T(i)T(j) }
        { (1-\alpha) T(m) + \alpha \sum_{i,j}\delta_{m,i+j}T(i)T(j) }.
        \label{eq:alpha_m}
\end{eqnarray}
The first equation defines the weight of sub-pileup events having multiplicities $i$ and $j$.
The second equation represents the pileup fraction at $m$th multiplicity bin.
True moments are then expressed recursively in terms of the measured moments at the lower multiplicity bins:
\begin{eqnarray}
	\ave{N^{r}}_{m}^{\rm t} &=& \cfrac{\ave{N^{r}}_{m}-\alpha_{m}\beta_{m}^{(r)}}{1-\alpha_{m}+2\alpha_m w_{m,0}}, \label{eq:final}
\end{eqnarray}
with 
\begin{eqnarray}
 \label{eq:final_corr}
	\beta_{m}^{(r)} &=& \mu_{m}^{(r)} + \sum_{i,j>0}\delta_{m,i+j}w_{i,j}\ave{N^{r}}_{i,j}^{\rm sub}, %\label{eq:final_corr}
\end{eqnarray}
and 
\begin{eqnarray}
  \mu_m^{(r)} =
  \begin{cases}
    \displaystyle
    2w_{m,0}\sum^{r-1}_{k=0}\binom{r}{k}\ave{N^{r-k}}_{0}^{\rm t}\ave{N^{k}}_{m}^{\rm t}
    & (m>0), \\
    \displaystyle
    \sum^{r-1}_{k=1}\binom{r}{k}\ave{N^{r-k}}_{0}^{\rm t}\ave{N^{k}}_{0}^{\rm t}
    & (m=0),
  \end{cases}
  \label{eq:mu_m}
\end{eqnarray}
where $\ave{N^{r}}^{t}_{m}$ and $\ave{N^{r}}_{m}$ represent $r$th order true and measured moments 
at $m$th multiplicity bin, respectively.
As can be seen from Eqs.~(\ref{eq:w_ij})--(\ref{eq:mu_m}), the necessary information to perform the 
pileup corrections are true multiplicity distribution $T(m)$ and pileup probability $\alpha$. 
In real experiments, we can only measure the multiplicity distributions 
including pileup events, and therefore, some models are needed to extract the true multiplicity distribution for single-collision events. 
One naive way is to use Glauber and particle production model to fit the multiplicity distribution as widely used for centrality determination. Details will be discussed in Sec.~\ref{sec:urqmd}.

%%%%%%%%%%%%%%%%%%%%%%%%%%%%%%%%%%%%%%%%%%%%%%%%%%%%%
\section{Pileup correction in UrQMD model\label{sec:urqmd}}
In this section, we demonstrate pileup correction method using UrQMD model. 
The package of version 3.4 is used and configured as cascade mode. Around 80 million events are generated with the impact parameter from 0~fm to 15~fm for Au+Au collisions at $\sqrt{s_{\rm NN}}=$3~GeV. Protons in rapidity and transverse momentum range of $-0.5<y<0$ and $0.4<p_{T}<2.0$ (GeV/c) are measured for cumulant calculations of proton multiplicity distributions. Collision centralities are defined by charged particle multiplicity at the pseudorapidity range of $|\eta|<1.0$, where protons and antiprotons are excluded to suppress self-correlations.

Cumulants up to the fourth order are calculated for two kinds of data sets. One is a raw UrQMD data for single-collision events. This will be a reference in the simulations without pileup events. The other one is a UrQMD data including both single-collision and pileup events. They will be referred to as "raw-UrQMD" and "pileup-UrQMD" for clarity.
In the simulation, the pileup events are generated via a superposition of two single-collision events. Two single-collision events in UrQMD are randomly added in terms of particle multiplicities with probability $\alpha$. Pileup-UrQMD samples are prepared for $0.1\%<\alpha<10\%$.

Pileup corrections are then performed for pileup-UrQMD data, with correction parameters determined in two ways as follows.
\begin{figure}[htbp]
	\begin{center}
	\includegraphics[width=0.5\textwidth]{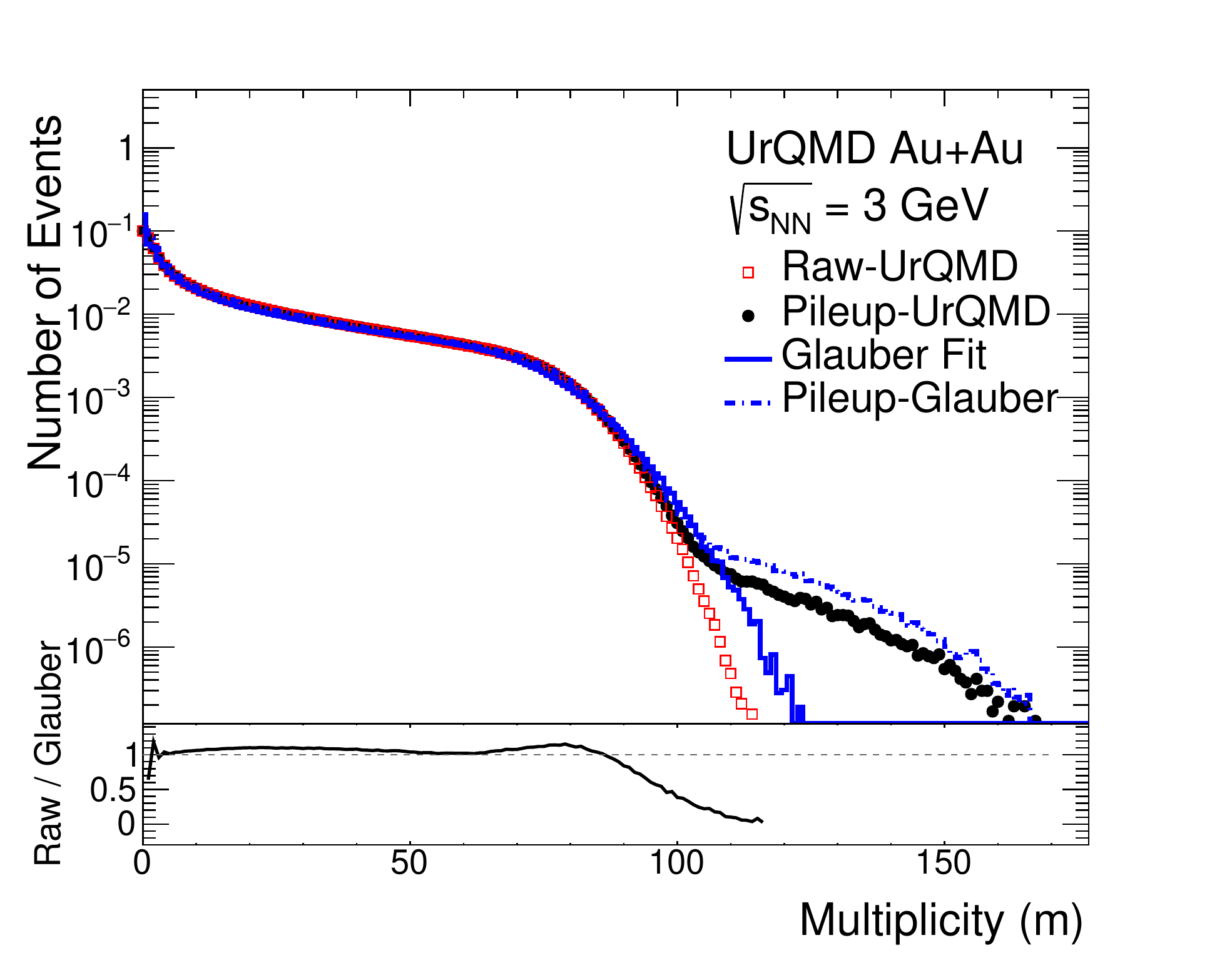}
	\end{center}
	\caption{
		(Color online) Charged particle multiplicity distribution in Au+Au collisions at $\sqrt{s_{\rm NN}}=$3~GeV from UrQMD model. The black circles are for pileup-UrQMD data with $\alpha=0.5$\%. The red squares show raw-UrQMD data for single-collision events, and a blue line is a Glauber fit to pileup-UrQMD data. The blue dashed line is pileup-Glauber distribution (pileup + single-collision) which is simulated using the single-collision distribution (blue line) extracted by Glauber fit.   
		}
	\label{fig:urqmd_fit}
\end{figure}
\begin{description}
\item[Case 1] \textbf{Closure test}\\ The raw-UrQMD data is used to determine the correction parameters according to Eqs.~(\ref{eq:w_ij}) and (\ref{eq:alpha_m}) with the know value of $\alpha$. This is a closure test as the same data sets are used to calculate cumulants and to determine correction parameters.
\item[Case 2] \textbf{Glauber fit} \\ The multiplicity distribution in pileup-UrQMD data is fitted by Glauber and particle production model.
In each event, we first determine the number of sources as:
\begin{equation}
    N_{\rm sc} = (1-x)N_{\rm part} + xN_{\rm coll},
\end{equation}
where $N_{\rm part}$ and $N_{\rm coll}$ represent the number of participant nucleons and binary collisions, respectively, given by the Glauber model. 
Final state multiplicity is then produced from $N_{\rm sc}$ according to negative binomial distributions given by
\begin{eqnarray}
	P_{\mu,k}(N) = \cfrac{\Gamma(N+k)}{\Gamma(N+1)\Gamma(k)} \cdot \cfrac{(\mu/k)^N}{(\mu/k+1)^{N+k}},
\end{eqnarray}
where $\mu$ is a mean value of particles generated from one source, and $k$ corresponds 
to an inverse of a width of the distribution.
The values of parameters employed are $\mu$ = 0.304, $k$ = 3.6 and $x $ = 0.11.
\end{description}

Figure~\ref{fig:urqmd_fit} shows multiplicity distribution in UrQMD model. 
In top panel red squares represent the raw-UrQMD data, and black circles are the pileup-UrQMD data with $\alpha=0.5$\%. A blue line is a Glauber fit to the pileup-UrQMD data. Parameters for pileup corrections can be constructed from blue and red lines in Fig.~\ref{fig:urqmd_fit} according to Eqs.~\ref{eq:w_ij} and \ref{eq:alpha_m} with known value of $\alpha$. In bottom panel of Figure~\ref{fig:urqmd_fit}, the black line represents a ratio of multiplicity distributions between raw-UrQMD data and Glauber fit. With the single-collision distribution extracted by Glauber fit, we then constructed pileup events and scan pileup fraction $\alpha$ to find a best fit to Pileup-UrQMD distribution. The best fit we found is shown as blue dashed line (pileup-Glauber) in Fig.~\ref{fig:urqmd_fit} with $\alpha=0.1$\% which is underestimated because of imperfect Glauber fit to pileup-UrQMD.  

\begin{figure*}[htbp]
	\begin{center}
	\includegraphics[width=0.8\textwidth]{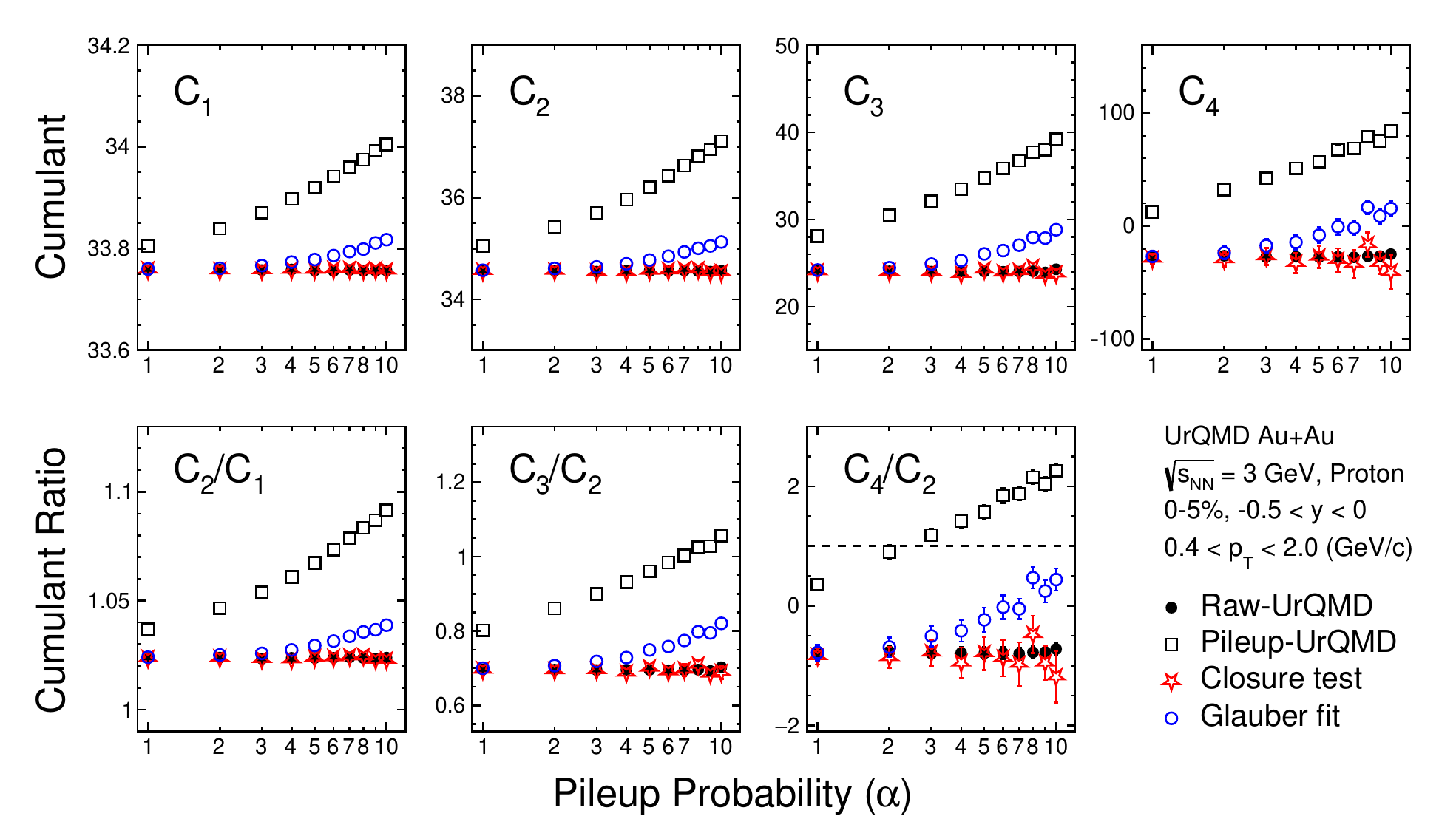}
	\end{center}
	\caption{
		(Color online) Proton cumulants and ratios up to fourth-order as a function of pileup level at top 5 \% in Au+Au collisions at $\sqrt{s_{NN}}$=3 GeV from UrQMD model. The black points show results from pure UrQMD model without any pileup simulation and the black squares are results from pileup simulated UrQMD data but are not pileup corrected. The red stars and blue circles are pileup corrected results which used true multiplicity distribution or Glauber fitted distribution as inputs respectively. 
		}
	\label{fig:urqmd_closure_glauber}
\end{figure*}

Figure~\ref{fig:urqmd_closure_glauber} shows cumulants and their ratios up to the fourth order as a function of pileup probability $\alpha$.
The black points are results for raw-UrQMD and black squares are for pileup-UrQMD data. 
It is found that the results of pileup-UrQMD deviate from raw-UrQMD with increasing the pileup probability. It is noteworthy that the value of $C_{4}/C_{2}$ changes sign from negative to positive then goes above unity with increasing the pileup probability. 
Results of pileup corrections for pileup-UrQMD data are shown in red stars for the closure test, and blue circles for the Glauber fit. It is found that the pileup correction works well for the wide range of pileup probability up to 10\%.
On the other hand, the pileup correction using the Glauber fit is not sufficient with $\alpha>1.0$\%. This would be because the fit quality is too poor to correct cumulants.
We have tried our best to tune the parameters of Glauber and two-component models to 
fit the multiplicity distribution in UrQMD data, but could not fit the distribution perfectly. 
Complicated processes of particle productions implemented in UrQMD may not be described by the two-component model.

%%%%%%%%%%%%%%%%%%%%%%%%%%%%%%%%%%%%%%%%%%%%%%%%%%%%%
\section{Unfolding\label{sec:unfolding}}
As discussed in Sec.~\ref{sec:urqmd}, 
pileup corrections depend on how one can precisely extract the true multiplicity distribution for single-collision events.
The issue is that the Glauber and particle production models, which are commonly used for centrality determination, cannot fit even the UrQMD data perfectly. 
A model independent way is necessary to make sure the quality of pileup corrections.

In this section, we propose a model independent way of an unfolding approach to determine the parameters for pileup corrections. The idea was originally developed to reconstruct the particle multiplicity distributions in terms of non-binomial detector efficiencies~\cite{Esumi:2020xdo}.
We found that the similar methodology is applicable in pileup corrections.
\begin{figure*}[htbp]
	\begin{center}
	\includegraphics[width=0.85\textwidth]{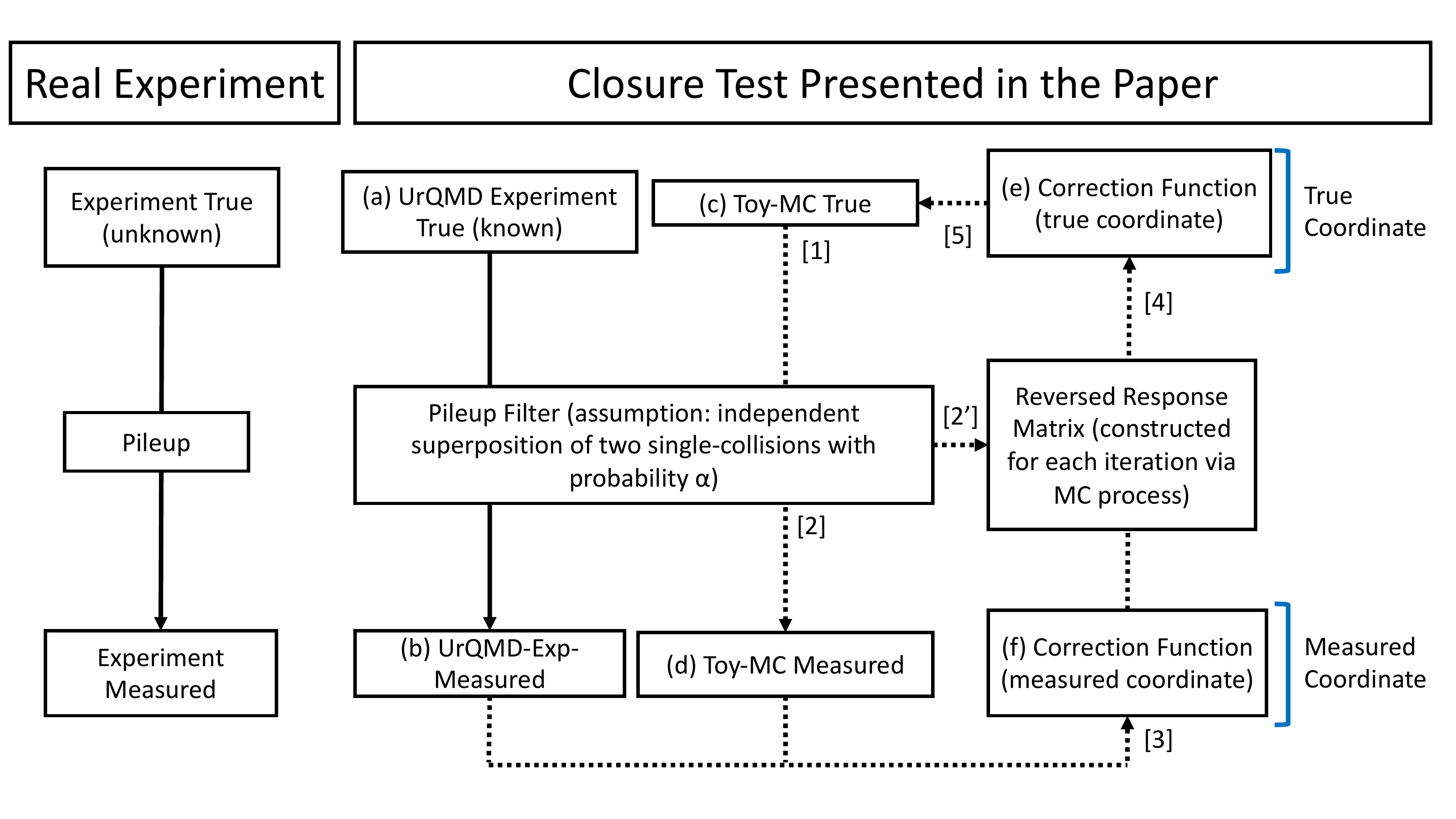}
	\end{center}
	\caption{
		Flowcharts in unfolding to extract the true multiplicity distribution. 
		The dotted arrows show the procedures repeated for iterations.
		}
	\label{fig:flowchart}
\end{figure*}

\begin{figure*}[htbp]
	\begin{center}
	\includegraphics[width=0.75\textwidth]{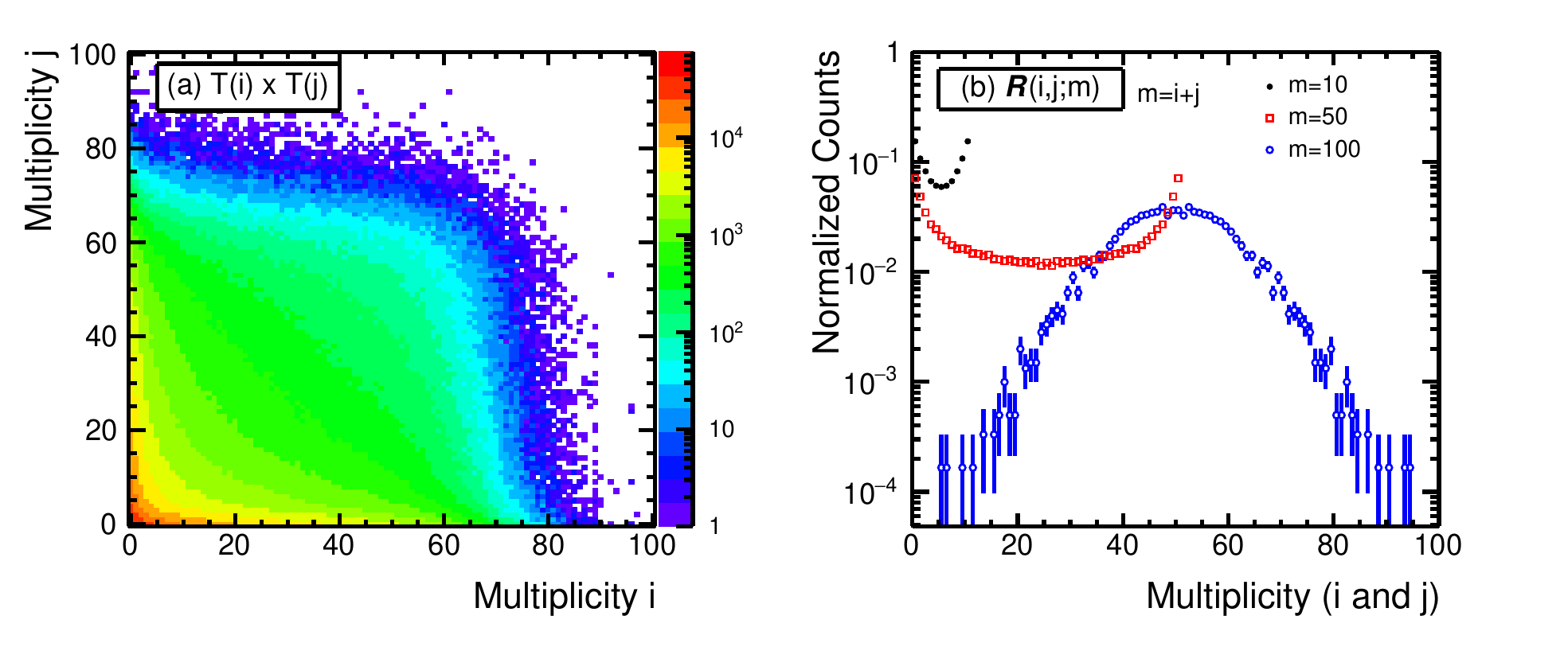}
	\end{center}
	\caption{
	(Left) Correlations between two independence multiplicity distribution in pileup events. (Right) Response matrices for 0th, 50th, and 100th iteration.
		}
	\label{fig:UrQMD_pileup_rm}
\end{figure*}

\begin{figure*}[htbp]
	\begin{center}
	\includegraphics[width=120mm]{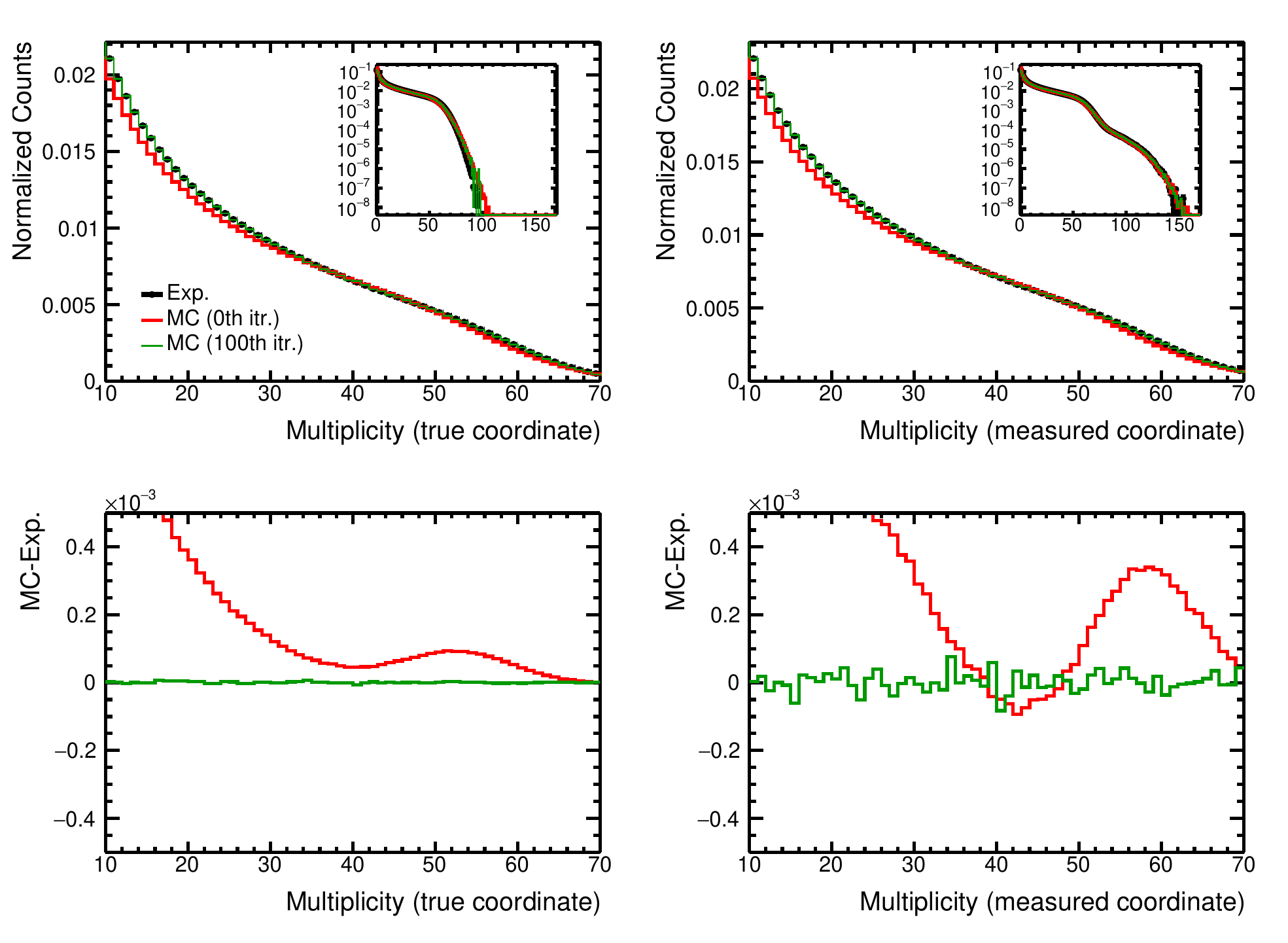}
	\end{center}
	\caption{
		(Top) Multiplicity distributions for UrQMD, 
		Glauber fit, and MC samples at 100th iteration.
		(Bottom) Difference between UrQMD and MC samples as a function of multiplicity. 
		Left-hand side panels are for the true coordinates, 
		while right-hand side panels are for the measured coordinates. The range of the x-axis is limited from 10 to 70 for illustration purpose.
		}
	\label{fig:UrQMD_pileup_dist}
\end{figure*}
Figure~\ref{fig:flowchart} depicts a flowchart for the unfolding procedure.
In real experiments, the multiplicity distribution is measured with 
the pileup events on top of the single-collision events as shown in 
the left row in Fig.~\ref{fig:flowchart}. "True" and "Measured" in the figure 
represent the multiplicity distributions for single-collisions 
and inclusive distributions for both single-collisions and pileup events. They are labeled as "(a) UrQMD experiment true" and "(b) UrQMD experiment measured", respectively. Both are related via a numerical process to generate pileup events called "Pileup Filter", which is defined as an independent superposition of two single-collision events with probability $\alpha$ for simplicity.
Similarly, we suppose Monte-Carlo samples labeled as "(c) toy-MC true" and "(d) toy-MC measured". They are also related via the same pileup filter between (a) and (b).
In the rest of this paper, samples in the top row will be referred to as "true coordinates", while the bottom row will be "measured coordinates".
The goal of the unfolding approach is to reconstruct (a) starting from (c). 
Detailed procedures are shown below: 
\begin{description}
    \item[0] Generate (a) UrQMD-experiment and (b) UrQMD-measured samples. They correspond to raw-UrQMD and pileup-UrQMD distributions in Sec.~\ref{sec:urqmd}.
	\item[1] Generate a (c) toy-MC distribution based on the Glauber model discussed in Sec.~\ref{sec:urqmd}. 
	\item[2] The pileup filter is applied to (c) to get (d) toy-MC measured distribution.
	\item[3] During the MC process from \textbf{1} to \textbf{2}, we compute the reversed response matrices, 
		${\cal R}$, numerically as shown in Fig.~\ref{fig:UrQMD_pileup_rm}. Note that any inversion procedure is not necessary here.
	\item[4] The correction function is determined by subtracting (b) from (d).
		It represents the difference between UrQMD-experiment and 
		toy-MC distributions in the measured coordinates. See lower panels in Fig.~\ref{fig:UrQMD_pileup_dist}.
	\item[5] The response matrix ${\cal R}$ is multiplied to (f) to get (e) the correction 
		functions in the true coordinates. 
	\item[6] By adding (e) to (c), the toy-MC distribution is modified to be closer to (a).
	\item[7] Repeat \textbf{1}--\textbf{6} until the correction functions become close enough to zero. 
\end{description}

The response matrices in \textbf{3} are defined as 
\begin{equation}
    T(i,j)=\sum_{i,j}\delta_{m;i,j}{\cal R}(i,j;m){\tilde T}(m),
\end{equation}
where ${\tilde T}(m)$ represents the probability distribution function of multiplicity in pileup events at measured coordinates, and $T(i,j)$ is a correlation between two multiplicities which forms pileup events.
$P(i,j)$ and {\cal R}(i,j;m) are shown in Fig.~\ref{fig:UrQMD_pileup_rm}. 
Distributions in Fig.~\ref{fig:UrQMD_pileup_rm}-(b) are projections of Fig.~\ref{fig:UrQMD_pileup_rm}-(a) onto a diagonal plane for $m=10$, $50$, and $100$ with $m=i+j$.
The response matrices relate the multiplicity $m$ observed in pileup events at measured coordinates and their original multiplicities from two single-collision events, $i$ and $j$ at true coordinates.
Note that the response matrices are determined during the numerical process of the pileup filter for each iteration.
Figure~\ref{fig:UrQMD_pileup_dist} shows multiplicity distributions and correction functions as a function of multiplicity
for true and measured coordinates, respectively. 
The initial distribution of MC samples are taken from the best fit of the Glauber model to the UrQMD-experiment distribution in the true coordinates. 
Nevertheless, there are large differences from the UrQMD-experiment distribution as can be seen in the correction functions.
After 100 iterations, the correction functions are found to be flat, which indicates that the multiplicity distribution for MC samples are successfully unfolded to UrQMD-experiment distributions. The resulting multiplicity distribution for the true coordinates can be used 
to determine the parameters for pileup corrections according to Eq.~\ref{eq:w_ij}.
\begin{figure*}[htbp]
	\begin{center}
	\includegraphics[width=0.85\textwidth]{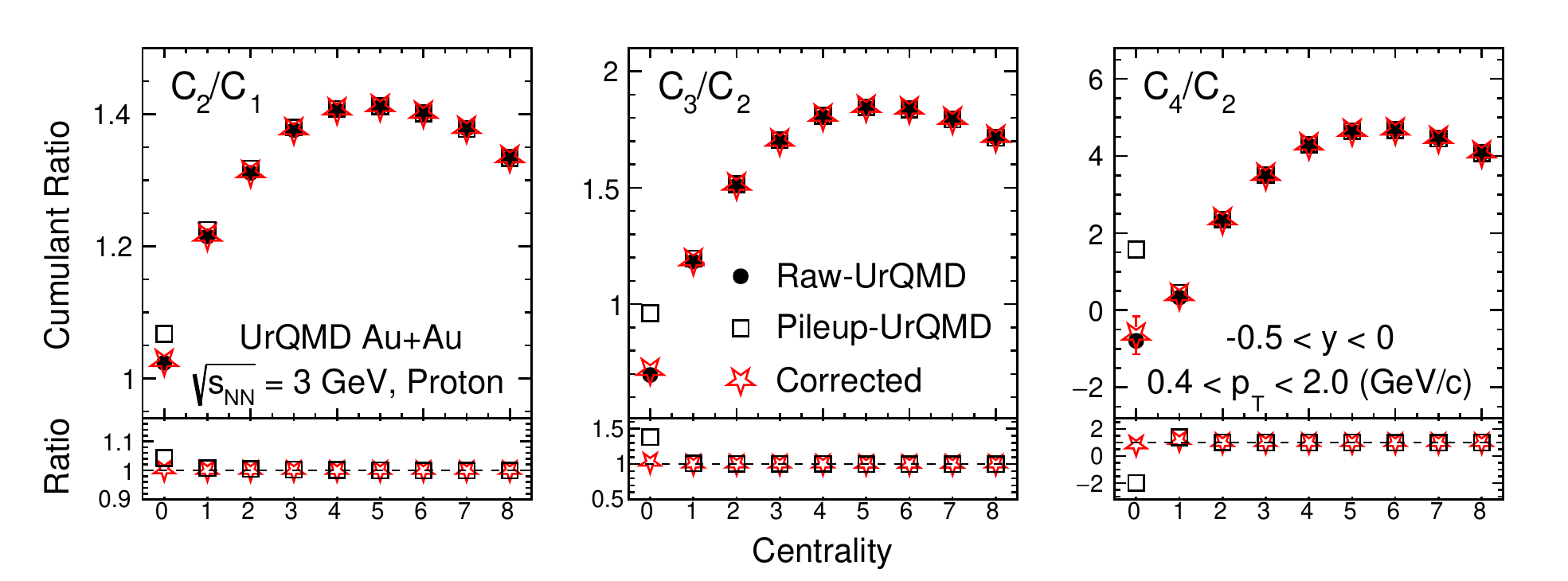}
	\end{center}
	\caption{(Color online) Centrality dependence of proton cumulant ratios up to fourth-order as a function of centrality in Au+Au collisions at $\sqrt{s_{\rm NN}}=$3~ GeV from UrQMD model. The label on x-axis from 0 to 8 represents 0-5\%, 5-10\%, 10-20\%, ..., 70-80\% centralities, respectively. The black points are results from raw-UrQMD model and black open squares are results from pileup-UrQMD data. The red stars in upper panels are pileup corrected results which utilized correction parameters extracted from the unfolding approach, and in lower panels the red stars and black squares represent ratios of corrected results and pileup-UrQMD results over raw-UrQMD results respectively.
		}
	\label{fig:urqmd_unfold}
\end{figure*}

In our simulations the preseted value of pileup probability $\alpha$ is used for the unfolding approach. In real experiments, the pileup probability can be basically calculated from the beam rates and thickness of the target material. To determine this more precisely, the unfolding approach needs to be repeated by varying the pileup probability to find the best parameter which yields the smallest values of $\chi^{2}/{\rm NDF}$.

Let us then move to the pileup corrections on cumulants. 
The multiplicity distribution for the true coordinate after 100 iterations 
is used to define the parameters for pileup corrections. 
Results are shown in Fig.~\ref{fig:urqmd_unfold} for up to the fourth-order cumulant 
as a function of centrality. 
Due to the effect of pileup events, the results at the most central collisions deviate from the true value of cumulants. The results of pileup correction using Glauber fits, however, still deviate from the true cumulants, which is because the Glauber fit is not perfect enough to describe the multiplicity distribution in UrQMD, as discussed in Sec.~\ref{sec:urqmd}.
We then apply pileup corrections with correction parameters determined by the unfolding approach. The results are consistent with true values of cumulants in the most central collisions. Therefore, it is concluded that our unfolding approach works well to determine the correction parameters for pileup corrections. In this work, we simulated pileup events by the superposition of two single-collision events, in fact the pileup events merged from more than two single-collision events can be also studied. In the unfolding approach the MC samples are taken from the best fit of the Glauber model to the UrQMD-experiment distributions. In principle, MC samples can start from any distributions like a flat distribution, but we propose to start from the distribution close to the experimental data to avoid possible systematics on the initial conditions of the MC samples.

\section{Summary\label{sec:summary}}
We perform detailed studies on the pileup correction for higher-order cumulants within UrQMD model. 
Two methods have been applied to determine the correction parameters and the results of pileup corrections are shown as a function of pileup probability. It is found that the pileup correction works well if the correction parameters are determined by the true multiplicity distribution of single-collision events and the known value of the pileup probability whereas the correction failed once the Glauber model fit is used to determine the parameters. This would be because the model of  particle  productions  are  different  from  those  implemented  in  UrQMD  model,  thus  the  fits  to  the  multiplicity distribution  does  not  work  perfectly.  The  study  indicates  that  the  pileup  correction  in  real  experiments  strongly depend on the model and the robustness of the pileup corrections are sensitive to the multiplicity distribution of single-collision events extracted from the measured multiplicity distributions.  To resolve the issue, we proposed an unfolding approach to extract the multiplicity distribution of single-collision events in a model independent way. The validity of this approach is verified in the UrQMD simulation, in which the pileup correction can successfully recover the true values of cumulants. Therefore, the model independent unfolding approach can be used in pileup corrections for the future measurement of higher-order cumulants in heavy-ion collisions. 

\section{Acknowledgement}
We thank D. Cebra, X. Dong, S. Esumi, S. Heppelmann and N. Xu for stimulating discussions. This work was supported by the National Key Research and Development Program of China (Grant No. 2020YFE0202002 and 2018YFE0205201), the National Natural Science Foundation of China (Grant No. 12122505, 11890711 and 11861131009), China Scholarship Council No. 201906770055 and JSPS KAKENHI Grant No. 19H05598.

\bibliography{main}% Produces the bibliography via BibTeX.

\end{document}